# Timing jitter characterization of free-running dual-comb laser with sub-attosecond resolution using optical heterodyne detection

SANDRO L. CAMENZIND,* DANIEL KOENEN, BENJAMIN WILLENBERG, JUSTINAS PUPEIKIS, CHRISTOPHER R. PHILLIPS, AND URSULA KELLER

*Department of Physics, Institute for Quantum Electronics, ETH Zurich, 8093 Zurich, Switzerland*
*casandro@phys.ethz.ch*

**Abstract:** Pulse trains emitted from dual-comb systems are designed to have low relative timing jitter, making them useful for many optical measurement techniques such as optical ranging and spectroscopy. However, the characterization of low-jitter dual-comb systems is challenging because it requires measurement techniques with high sensitivity. Motivated by this challenge, we developed a technique based on an optical heterodyne detection approach for measuring the relative timing jitter of two pulse trains. The method is suitable for dual-comb systems with essentially any repetition rate difference. Furthermore, the proposed approach allows for continuous and precise tracking of the sampling rate. To demonstrate the technique, we perform a detailed characterization of a single-mode-diode pumped Yb:CaF$_2$ dual-comb laser from a free-running polarization-multiplexed cavity. This new laser produces 115-fs pulses at 160 MHz repetition rate, with 130 mW of average power in each comb. The detection noise floor for the relative timing jitter between the two pulse trains reaches $8.0\times10^{-7}$ fs$^2$/Hz (~ 896 zs/$\sqrt{\text{Hz}}$), and the relative root mean square (rms) timing jitter is 13 fs when integrating from 100 Hz to 1 MHz. This performance indicates that the demonstrated laser is highly compatible with practical dual-comb spectroscopy, ranging, and sampling applications. Furthermore, our results show that the relative timing noise measurement technique can characterize dual-comb systems operating in free-running mode or with finite repetition rate differences while providing a sub-attosecond resolution, which was not feasible with any other approach before.

## 1. Introduction

A dual-modelocked laser, in a most general sense, means that two modelocked pulse trains can be generated from a single laser cavity. Depending on the application, these two pulse trains are mainly distinguished by their center wavelength and/or their pulse repetition rate. Synchronized dual-color ultrafast lasers have been explored extensively for pump-probe measurements with a tunable probe pulse, originally using dye lasers in the 1970s [1] and then Ti:sapphire lasers in the 1990s [2, 3]. The frequency comb revolution [4-6] and dual-comb spectroscopy [7, 8] motivate dual-comb measurements with two pulse trains with different pulse repetition rates for equivalent time sampling [9-12]. Initially such dual-comb measurements were based on two separate frequency comb sources that had to be mutually stabilized with four stabilization loops for their comb spacing and offset [13-17]. Dual-comb modelocking from a single laser cavity with different pulse repetition rates have been first demonstrated with polarization duplexing [18], and in ring lasers with circular direction duplexing [19, 20]. This resulted in a paradigm shift for dual-comb spectroscopy using only a free-running dual-comb modelocked laser [21]. Dual-comb measurements have become a hot topic as demonstrated with recent review articles [8, 22-26] and many different application demonstrations such as precise optical ranging [27-32], high-speed pump-probe measurements via equivalent time sampling [33-36], electro-optic sampling spectroscopy [37], high-resolution time-domain spectroscopy [38, 39], etc.

To make dual-combs suitable for applications, it is crucial to ensure that the two pulse trains exhibit low relative timing jitter (i.e., the uncorrelated timing jitter between the two repetition rates) for the duration of a measurement. Minimizing the timing jitter is important for resolving the optical comb lines in dual-comb spectroscopy, and for achieving a high temporal resolution in equivalent time sampling systems [40]. Nonetheless, measuring this relative jitter with sufficient accuracy is challenging due to the high sensitivity required for state-of-the-art low-jitter dual-comb systems. As the rapid advance of dual-comb technology opens the door for industrial and scientific applications in various fields, it is of high interest to find techniques that allow for accurate and reliable measurements of the relative timing jitter. Such measurements are particularly important for free-running dual-combs, which have been extensively explored due to their reduced complexity compared with fully stabilized systems [18-20, 41-50].

Since many dual-comb measurements are influenced by sub-femtosecond jitter in the relative optical delay between the combs [51-60], very sensitive noise measurements are needed. This jitter arises due to uncorrelated fluctuations in the repetition rates of the two combs and is therefore referred to as uncorrelated timing jitter. Here, our goal is to develop a method for measuring the noise of free-running dual-combs having (almost) arbitrary repetition rate and repetition rate difference, while resolving sub-femtosecond timing jitter over fast time scales. Our method is based on heterodyne beat notes between the combs and a pair of continuous wave lasers. To experimentally demonstrate this method, we first present a new low-noise dual-comb laser system, and then use the new noise measurement method to characterize the laser's timing jitter with high precision. In order to motivate the need for a new noise characterization approach and put our measurement method into context, in the following paragraphs we discuss the most relevant techniques and their trade-offs.

There already exist promising methods for estimating the uncorrelated timing jitter. An appealing solution is provided by Modsching *et al.* [60]. By using a fast photodiode, they detect the repetition rate of each comb and the corresponding harmonics, up to harmonic number $N > 100$. By electronic mixing, they obtain the frequency difference between the $103^{rd}$ harmonic of the two repetition rates, i.e., $103 \cdot \Delta f_{rep}$. The resulting signal is analyzed on a commercial phase-noise analyzer that relies on an electronic reference oscillator [60]. However, comparing the detected difference frequency signal to the timing reference of the spectrum analyzer sets an inherent lower boundary for the measurement sensitivity since the phase noise of the timing reference needs to be lower than the signal noise. Furthermore, this approach is only suitable for dual-comb systems with high repetition rate difference $\Delta f_{rep}$, which might not always be the desired setting of a dual-comb, depending on the specific application. A closely related approach suitable for assessing the long-term stability of repetition rate difference relies on the digital retrieval of the difference frequency signal from the 100th harmonic of the two repetition rates recorded as spectrograms [50].

Another class of techniques is designed to measure the relative timing jitter between a pair of lasers that are loosely locked together with active feedback on the repetition rate of one of the lasers. In these techniques, $\Delta f_{rep}$ is nominally zero and the relative delay between the pulses is always much smaller than the inverse of the laser repetition rate. In the approach of [61] belonging to this family of techniques, the repetition rate of each comb is measured with separate photodiodes. The phase difference between these signals is determined electronically and is locked to $\pi/2$. Timing jitter then leads to deviations in the phase of this electronic signal from $\pi/2$. Higher precision can be obtained by the combination of two phase detectors, and operation on different harmonics of the photodiode signal [51], but the signal is still limited by the speed of photodetectors and electronics.

Even higher resolution can be achieved by using optical techniques in case of femtosecond optical pulses. The approach based on a balanced optical cross-correlator (BOC) relies on intensity cross-correlation in a nonlinear crystal by sum-frequency generation [52, 54-58]. With BOC, it is possible to measure timing jitter with a resolution below 10 zs/$\sqrt{Hz}$, and in addition, the detection approach is free of excess amplitude-noise to phase-noise coupling. However, the approach has the inherent limitation that timing errors should be smaller than the pulse duration, and hence requires setting $\Delta f_{rep} = 0$ which is not appropriate for some dual-comb systems, especially free-running dual-combs.

In contrast to non-collinear intensity cross-correlation between two pulse trains applied in a BOC, an interferometric cross-correlation measurement can provide higher sensitivity for timing jitter measurements due to the ultrahigh coherence between longitudinal modes of mode-locked lasers as demonstrated by Chen *et al.* [53]. To obtain a phase discrimination signal, they superimposed the two pulse trains and detected a narrow spectral band at each extreme of the optical spectrum with a photodetector. Subsequently, the beat note signal from the optical beat detection was shifted down to baseband by electronic mixing. The signal-to-noise ratio (SNR) was limited because the measurement did not use the total optical power of the pulse trains, which impacts the measurement resolution. To overcome this limitation, Hou *et al*. demonstrated more broadband detection to better use the optical power. They obtained a shot-noise-limited discrimination signal close to a single cycle [59]. However, since this powerful approach still falls into the category of noise measurements that require the lasers to be loosely locked, it is not ideal for noise measurements on e.g. free-running dual-comb lasers.

Another option for characterizing the timing jitter of a dual-comb system that does not rely on locking the repetition rates of the laser is the indirect phase comparison of two photodetector outputs [62]. An electronic reference oscillator is mixed with the repetition rate of each comb, so that the noise of the reference oscillator is common to both of the output signals. Hence, upon taking the difference frequency signal, the influence of the reference oscillator cancels. This method provides a highly suitable approach for measuring the uncorrelated timing jitter of two pulse trains as the method works for (almost) any repetition rate difference while keeping amplitude-noise to phase-noise conversion at low levels, thereby rendering this phase noise measurement less sensitive to intensity noise.

However, a limitation of electronic measurements of $\Delta f_{rep}$ is that they are limited to tens of gigahertz signals, which implies that only a low-order harmonic of $\Delta f_{rep}$ (typically on the order of $10^2$) can be detected. Increased sensitivity can be achieved by utilizing optical frequencies instead of electronic ones since then a high-order harmonic of $\Delta f_{rep}$ (typically on the order of $10^5$) can be measured. In the technique of adaptive dual-comb spectroscopy by Ideguchi *et al*. [63], a pair of cw lasers are mixed with the optical frequency combs, enabling real-time tracking of a high-order harmonic of $\Delta f_{rep}$.

Here, we present a multiheterodyne detection approach based on similar beat note signals as in [63]. However, instead of deploying the measured signals in the context of dual-comb spectroscopy, we use them for measuring and analyzing the relative timing jitter of two pulse trains with high sensitivity. The technique is suitable for analyzing the noise of both free-running and stabilized dual-comb systems, provided the repetition rate difference is sufficiently small compared to the repetition rate so that cross-correlation signals between the two combs can be resolved. Additionally, it provides the repetition rate difference continuously as a function of time which can be used for dual-comb signal correction with methods like adaptive sampling either in post-processing or with an active feedback loop [63, 64].

Using this detection approach, we demonstrate the first detailed characterization of the uncorrelated repetition rate noise of a free-running femtosecond dual-comb laser. This laser is the first single-mode-diode pumped dual-comb laser using the polarization multiplexing technique introduced in 2015 with an additional intracavity birefringent crystal [18]. This is in contrast of using an intrinsic polarization-multiplexing approach inside an optically anisotropic gain crystal as shown more recently in 2021 [50]. The demonstrated laser has an average output power of 130 mW per comb, a pulse duration of 115 fs, a pulse repetition rate of 160 MHz, and a small repetition rate difference of $\Delta f_{rep}$ = 1.286 kHz. It is constructed in a robust prototype housing for low noise operation and is passively cooled. With the new noise characterization technique, we measured a low relative root mean square (rms) timing jitter of 13 fs for the frequency interval [100 Hz, 1 MHz] with the detection noise floor at $8.0 \times 10^{-7}$ fs$^2$/Hz. The approach furthermore allows to effectively separate phase noise from intensity noise while keeping amplitude-to-phase noise coupling minimal.

This paper is structured as follows: The proposed timing jitter measurement technique based on multiheterodyne detection is described in section 2. In section 3, we discuss the practical implementation. In section 4 we present the free-running dual-comb laser system and apply the multiheterodyne detection technique for characterizing its timing jitter. In section 5, the technique is compared to a noise measurement based on the interferograms generated from interference between the two combs, without any cw laser, to show its potential and validate the approach. Finally, in section 6 we discuss and conclude the work.

## 2. Optical heterodyne detection of repetition rate difference for noise analysis

In this section, we will discuss the basic concept of our noise measurement technique. The proposed measurement technique can be used to determine the relative timing jitter of two modelocked lasers that form a dual-comb system. A dual-comb consists of two combs with slightly different characteristic frequencies for the individual comb lines:

$$f_N^{(comb\text{-}i)} = f_{CEO}^{(comb\text{-}i)} + N^{(comb\text{-}i)} \cdot f_{rep}^{(comb\text{-}i)} \quad (1)$$

with $f_{rep}^{(comb\text{-}i)}$ being the repetition rate frequency and $f_{CEO}^{(comb\text{-}i)}$ the carrier-envelope offset (CEO) frequency [4] of comb $i \in \{1,2\}$. According to Eq. (1) the contribution of the repetition rate fluctuations to the fluctuations of the $N$-th comb line scales with the comb line index $N^{(comb\text{-}i)}$. An optical frequency measurement is thus beneficial as it implicitly involves an enhancement of the fluctuations of the pulse repetition rate by the comb line index, so that small frequency fluctuations become easier to measure.

To determine the uncorrelated timing jitter of the two combs, we are thus generating beat note signals between each comb and a pair of narrow-linewidth single-frequency lasers with frequencies $f_{(cw\text{-}1)}$ and $f_{(cw\text{-}2)}$ which results in four beat note signals with frequencies

$$f_{beat,(cw\text{-}j)}^{(comb\text{-}i)} = f_{CEO}^{(comb\text{-}i)} + N_{(cw\text{-}j)}^{(comb\text{-}i)} \cdot f_{rep}^{(comb\text{-}i)} - f_{(cw\text{-}j)} \quad (2)$$

of comb $i \in \{1,2\}$ with cw laser $j \in \{1,2\}$ as visualized in Fig. 1. Because in the actual experiment we can only measure signals with positive frequencies, the frequency of the detected signal is $\left|f_{beat,(cw\text{-}j)}^{(comb\text{-}i)}\right|$. The comb line indices $N_{(cw\text{-}j)}^{(comb\text{-}i)}$ are such that $f_{meas,(cw\text{-}j)}^{(comb\text{-}i)}$ is always lower than $f_{rep}^{(comb\text{-}i)}/2$, meaning that for each comb we consider the beat notes of the cw lasers with the closest comb-line.

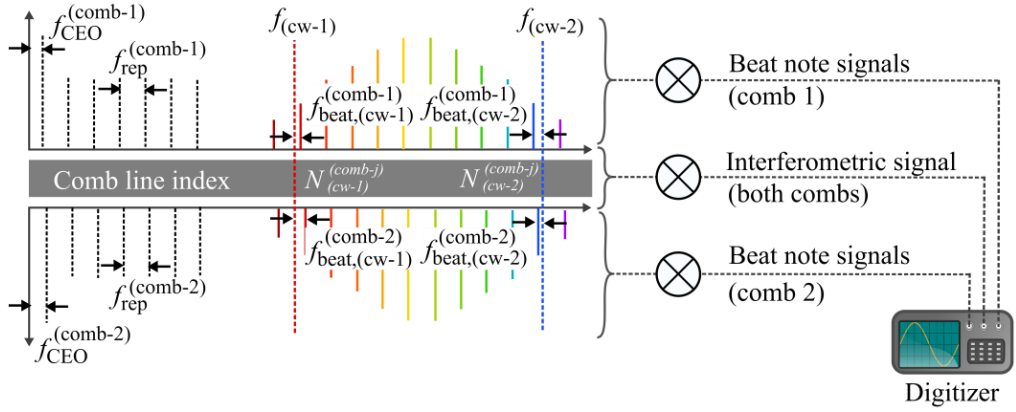

**Fig. 1.** Schematic of the measurement setup based on an optical heterodyne detection approach for measuring the relative timing jitter of two pulse trains. The two outputs of the dual-comb system beat with a pair of narrow-linewidth continuous-wave (cw) lasers on a photodetector. The signal is subsequently filtered at half of the repetition rate with a low-pass filter (LPF). Simultaneously, an interferometric signal is obtained through optical beat detection of the two frequency combs, followed by an identical LPF at half of the repetition rate. The three electronic signals are digitized simultaneously on an oscilloscope.

By taking the difference between the beat note frequencies corresponding to the same cw laser $\Delta f_{beat,(cw\text{-}j)} \equiv f_{beat,(cw\text{-}j)}^{(2)} - f_{beat,(cw\text{-}j)}^{(1)}$, the frequency $f_{(cw\text{-}j)}$ cancels which results in a signal that depends only on the comb properties. This implies that the frequency noise properties of the cw lasers are not critical for the measurement itself. In summary the quantity $\Delta f_{beat,(cw\text{-}j)}$ is insensitive to the relative fluctuations of the two cw lasers:

$$\begin{aligned} \Delta f_{beat,(cw\text{-}j)} &\equiv f_{beat,(cw\text{-}j)}^{(comb\text{-}2)} - f_{beat,(cw\text{-}j)}^{(comb\text{-}1)} \\ &= \Delta f_{CEO} + N_{(cw\text{-}j)}^{(comb\text{-}1)} \cdot \Delta f_{rep} + \left(N_{(cw\text{-}j)}^{(comb\text{-}2)} - N_{(cw\text{-}j)}^{(comb\text{-}1)}\right) \cdot f_{rep}^{(comb\text{-}2)}. \end{aligned} \quad (3)$$

To isolate the relative fluctuations of the carrier-envelope offset frequencies $f_{CEO}$, thereby rendering the measurement insensitive to fluctuations of $f_{CEO}$, we can take the difference between the expressions for $\Delta f_{beat,(cw-j)}$:

$$\begin{aligned} f_{signal} &= \Delta f_{beat,(cw-2)} - \Delta f_{beat,(cw-1)} \\ &= \left( N_{(cw-2)}^{(comb-1)} - N_{(cw-1)}^{(comb-1)} \right) \cdot \Delta f_{rep} \\ &\quad + \left[ \left( N_{(cw-2)}^{(comb-2)} - N_{(cw-1)}^{(comb-2)} \right) - \left( N_{(cw-2)}^{(comb-1)} - N_{(cw-1)}^{(comb-1)} \right) \right] \cdot f_{rep}^{(comb-2)} \\ &= \Delta N \cdot \Delta f_{rep} + \Delta n \cdot f_{rep}^{(comb-2)} \end{aligned} \quad (4)$$

where $\Delta N = \left( N_{(cw-2)}^{(comb-1)} - N_{(cw-1)}^{(comb-1)} \right)$ is the number of optical comb lines between the two cw lasers for comb 1, and $\Delta n \equiv \left[ \left( N_{(cw-2)}^{(comb-2)} - N_{(cw-1)}^{(comb-2)} \right) - \left( N_{(cw-2)}^{(comb-1)} - N_{(cw-1)}^{(comb-1)} \right) \right]$ defines the difference between $\Delta N$ and the number of optical comb lines between the two cw lasers counted with the spacing of comb 2. Hence, this double-difference signal is only sensitive to changes in the pulse repetition rates. For small values of the repetition rate difference we may assume that $\Delta n = 0$, so that

$$f_{signal} = \Delta N \cdot \Delta f_{rep}. \quad (5)$$

For $\Delta n \neq 0$, which only happens for large $\Delta f_{rep}$, we can infer $f_{rep}^{(comb-2)}$ from the beat note measurement itself or from a separate measurement with a microwave spectrum analyzer (MSA). We then simply redefine $f_{signal}$ as

$$\begin{aligned} f_{signal} &= \left[ \Delta f_{beat,(cw-2)} - \Delta f_{beat,(cw-1)} \right] - \Delta n \cdot f_{rep}^{(comb-2)} \\ &= \Delta N \cdot \Delta f_{rep}, \end{aligned} \quad (6)$$

which leads to the same final result.

To determine the repetition rate difference of the dual-comb system, we divide $f_{signal}$ by the comb line index difference $\Delta N$. If the cw laser frequencies are located on either side of the optical spectrum, the comb line index difference is determined by the optical bandwidth divided by the repetition rate. Depending on the parameters of the specific lasers, $\Delta N$ can reach values $> 10^5$.

To infer the comb line index difference $\Delta N$, we use the interferometric signal between the two frequency combs. Specifically, $\Delta f_{rep} \approx 1/\Delta T$, where $\Delta T$ is the delay between adjacent peaks of the envelope of the interferograms that repeat every $1/\Delta f_{rep}$ when the pulses from the two pulse trains coincide. The envelope of each interferogram is reconstructed with the absolute value of the analytic signal of the interferograms which is extracted through a Hilbert transform [32]. Even though this measurement of $\Delta T$ from the interferograms involves coarse sampling, it is sufficient to obtain an accurate estimate of the comb line index difference:

$$\Delta N = \text{round}\left( f_{signal} \cdot \Delta T \right), \quad (7)$$

as can be seen from Eq. (5). By following this procedure, we find

$$\Delta f_{rep}(t) = \frac{f_{signal}(t)}{\Delta N}, \quad (8)$$

which can be used to deduce the phase noise power spectral density. Note that deriving $\Delta f_{rep}(t)$ in this way significantly increases the sensitivity over a direct radio-frequency (RF) measurement of the repetition rate difference. Aside from finding the phase noise power spectral density, the repetition rate $\Delta f_{rep}(t)$ as a function of time also helps to improve the performance of a dual-comb system with methods like adaptive sampling either in post-processing or with an active feedback loop [63, 64]. Furthermore, it could be used to infer $\Delta f_{CEO}(t)$ by utilizing Eq. (3) together with the absolute comb line indices corresponding to one of the cw lasers. To find the absolute comb line indices one would need to perform an accurate measurement of the cw laser's frequency, together with the repetition rate of the frequency combs. Alternatively, the beat note signal from Eq. (3) corresponds to a particular RF comb line, and hence allows for a direct relative coherence assessment [20, 65]. For example, its noise spectrum can be linked to the full-width half maximum (FWHM) linewidth of a single RF comb tooth via the β-separation line approximation [66].

The information of $\Delta f_{rep}(t)$ as a function of time – with the sampling rate only limited by the data acquisition – can now be used for finding the phase noise power spectral density (PN-PSD) [61, 67, 68]

$$S_\varphi(f) = 2 \cdot \frac{\left| F\left\{ \Delta f_{rep}(t) - \langle \Delta f_{rep}(t) \rangle \right\} \right|^2}{f^2}, \qquad (9)$$

where the factor of two indicates that we consider the one-sided power spectral density. To be consistent with the definition, we will exclusively work with one-sided power spectral densities. In Eq. (9) we were using the mean of the repetition rate difference $\langle \Delta f_{rep}(t) \rangle$ as well as the Fourier transform

$$X(f) = F\{x(t)\} = \int_{-\infty}^{\infty} x(t) \cdot e^{-2\pi i f t} dt. \qquad (10)$$

According to Parseval's theorem, the PN-PSD can be used for computing the rms jitter of the relative timing between the two pulse trains

$$\tau_{[f_1, f_2]}^{rms} = \frac{\sqrt{\int_{f_1}^{f_2} S_\varphi(f) df}}{2\pi \cdot \Delta f_{rep}} \cdot \frac{\Delta f_{rep}}{f_{rep}} = \frac{\sqrt{\int_{f_1}^{f_2} S_\varphi(f) df}}{2\pi \cdot f_{rep}}. \qquad (11)$$

In the following we will refer to this type of jitter as relative rms timing jitter between the two pulse trains. The comb factor $\Delta f_{rep}/f_{rep}$ was used to relate the RF-domain with the optical domain based on the average repetition rate and repetition rate difference. Such rms timing jitter values capture the fluctuations between the actual delay and the delay that would occur in the absence of jitter.

Another useful quantity to characterize the noise is the integrated period timing jitter. This quantity is defined as the difference between the time-varying period of the signal being measured and the average period. This quantity provides information about the variation in the period of a nearly-periodic signal. The integrated period timing jitter can be calculated as

$$\tau_{[f_1, f_2]}^{period} = \frac{\sqrt{\int_{f_1}^{f_2} \chi(f) \cdot S_\varphi(f) df}}{2\pi \cdot \Delta f_{rep}} \cdot \frac{\Delta f_{rep}}{f_{rep}}$$

$$= \frac{\sqrt{\int_{f_1}^{f_2} \chi(f) \cdot S_\varphi(f) df}}{2\pi \cdot f_{rep}}, \qquad (12)$$

with

$$\chi(f) = \left| 1 - \exp\left( -2\pi i \cdot \frac{f}{\Delta f_{rep}} \right) \right|^2 \qquad (13)$$

being the filter function, which is periodic in frequency with vanishing amplitude at integer multiples of $\Delta f_{rep}$, i.e., it suppresses noise around DC and all harmonics of $\Delta f_{rep}$ [69]. The period timing jitter is relevant for triggered data acquisition as applied e.g. for equivalent time sampling measurements or coherent averaging in dual-comb spectroscopy because the repeating trigger signal corrects for long-term timing drifts.

## 3. Experimental implementation

### 3.1 Experimental setup

In this subsection, we discuss the experimental setup used for measuring the uncorrelated timing jitter according to the procedure described in section 2. To reduce the complexity of the measurement system, we decided to implement it as a fully polarization-maintaining (PM) fiber-based breadboard setup. Consequently, for each modelocked laser emitted by the dual-comb system we couple the light into an optical fiber splitter as depicted schematically in Fig. 2. Similarly, each of the two cw lasers (CTL 1050, Toptica Photonics) are coupled into one of the two input ports of a 50:50 fiber splitter. Each output of this splitter is then combined with one of the two combs and detected with a reverse-biased 0.8-mm² InGaAs photodiode. The superposition of the two cw lasers with each frequency comb leads to the generation of two time traces that encode together four beat frequencies $f_{\text{beat,(cw-j)}}^{(\text{comb-i})}$, $i, j \in \{1, 2\}$ as visualized in Fig. 2. The cw lasers are tuned to $f_{(\text{cw-1})} = 277.66$ THz and $f_{(\text{cw-2})} = 286.14$ THz. Additionally, we combine the remaining output ports of the optical fiber splitter of each comb to generate an interferometric signal on a third InGaAs photodiode (DET08CFC, Thorlabs Inc.). The time-domain interferometric signal exhibits a periodic succession of interferograms that repeat every $1/\Delta f_{\text{rep}}$ when the pulses from the two modelocked lasers coincide temporally. As the relevant part of the RF-spectrum spans the frequency range from DC to $f_{\text{rep}}/2$, the detected signals are low-pass filtered at the Nyquist frequency $f_{\text{rep}}/2$ and digitalized simultaneously on an oscilloscope with 12-bit vertical resolution (WavePro 254HD, Teledyne LeCroy).

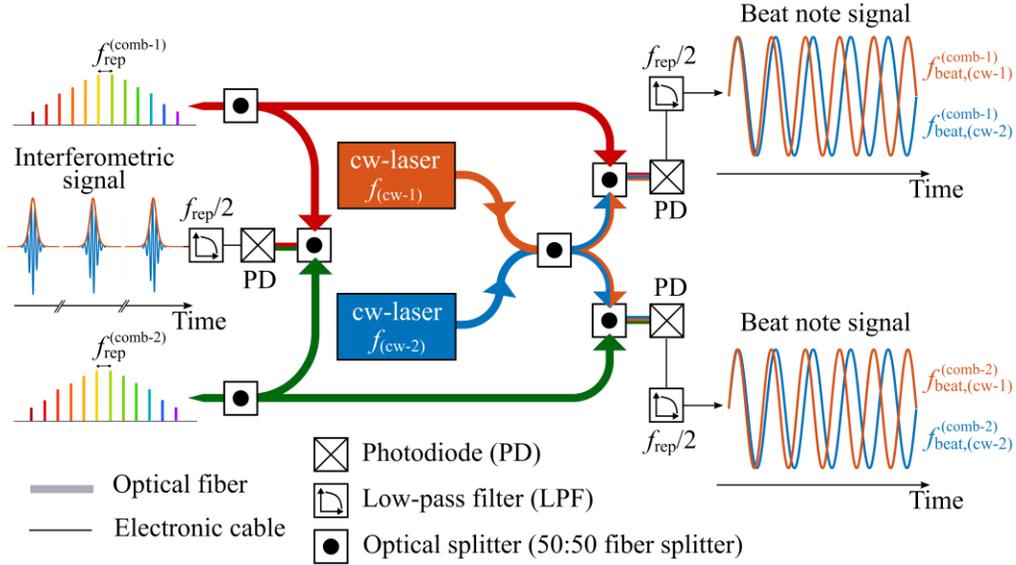

**Fig. 2.** The two outputs of the dual-comb system with small $\Delta f_{\text{rep}}$ are combined with a 50:50 fiber splitter. Each comb simultaneously beats with two independent cw lasers to isolate a single comb line. Simultaneously, the two combs are combined to generate an interferometric signal. The beat note signals, as well as the interferometric signal, are detected on a photodiode (PD) and low-pass filtered at $f_{\text{rep}}/2$.

It is important to mention that the recorded signals are affected by the phase noise of the cw lasers, which can be larger than the fluctuations of the optical comb lines. However, since both channels share the cw lasers and thus the corresponding phase noise, this noise affects both channels equally. Consequently, the noise inherent to the cw lasers cancels in the post-processing as discussed in section 3.2.

*3.2 Post-processing and analysis*

In this subsection, we introduce the analysis routine used to extract the noise characteristics inherent to a dual-comb system from the data recorded with the setup shown in Fig. 2. The experimental implementation of the noise measurement described in section 2 consists of the simultaneous recording of three time traces. Notice that before performing the fast Fourier transform (FFT) algorithm on a time trace, it is crucial to apply an appropriate window function to avoid leakage phenomena [70] which can be an issue in situations where the power spectral density diverges at zero frequency [71].

The first step of the analysis routine is to transform the two beat note signals to the frequency domain using an FFT as visualized in Fig. 3. The Fourier-transformed data of the two beat note signals each contain two peaks below $\Delta f_{\text{rep}}/2$, one for each cw laser beating with the closest optical comb line. Those peaks in the frequency domain contain information about the frequency fluctuations of the comb lines. To extract this information, the second step is to determine the positions of each of those peaks with a first-order moment integral of the magnitude squared of the individual peaks. We then apply a digital bandpass filter to select only the signal corresponding to the individual beat note frequencies $f_{\text{beat,(cw-j)}}^{(\text{comb-i})}$ associated with the beating of frequency comb $i \in \{1,2\}$ with cw laser $j \in \{1,2\}$. In a third step, we then apply an inverse fast Fourier transform (iFFT) to go back to the time domain which yields four isolated beat note signals. The instantaneous frequencies of beat notes produced by the interaction with the same cw laser are visually similar (see Fig. 3(g),(h)) as expected for high relative coherence between the combs. By digitally mixing each pair of complex signals corresponding to the same cw laser, we obtain two signals at the difference frequency $\Delta f_{\text{beat,(cw-j)}}(t)$ according to Eq. (3). Note that by mixing the beat note signals, the frequency fluctuations inherent to the cw lasers cancel as discussed in section 2. Consequently, the carrier frequency $\Delta f_{\text{beat,(cw-j)}}(t)$ of the resulting signal is more stable than the frequencies $f_{\text{beat,(cw-j)}}^{(\text{comb-i})}$ of the individual beat notes themselves. To isolate background noise, we then apply a digital bandpass filter with a 3-dB bandwidth of 3 MHz and centered around $\Delta f_{\text{beat,(cw-j)}}(t)$. In a final step, the two isolated signals with frequencies $\Delta f_{\text{beat,(cw-j)}}(t), j \in \{1,2\}$ are digitally mixed and used together with the interferometric data to infer $\Delta f_{\text{rep}}(t)$ as described by Eqs. (6)-(8). To extract the instantaneous frequency of the mixing product, we took the derivative of the instantaneous phase after unwrapping. Alternative methods developed for instantaneous frequency estimation could also be used for this purpose.

As discussed in section 2, it is possible to extract the phase noise power spectral density (PN-PSD) from $\Delta f_{\text{rep}}(t)$ by using a Fourier transform and Eq. (9). However, the discrete Fourier transform – which is the preferred approach to find the PN-PSD computationally – has some subtleties. The naïve approach based on multiplying the time trace with a suitable window function, performing an FFT, and calibrating the results with the rms of the window function to guarantee energy conservation, will typically lead to significant fluctuations in the spectrum due to the lack of spectral averaging. A possible alternative to avoid this issue is to slice the time trace into several smaller overlapping segments. The length of the individual segments is chosen such that it is an integer multiple of $1/\Delta f_{\text{rep}}$. For each segment, we then apply a Hanning window prior to the FFT. This approach is sometimes also referred to as Welch's overlapped segmented average [72].

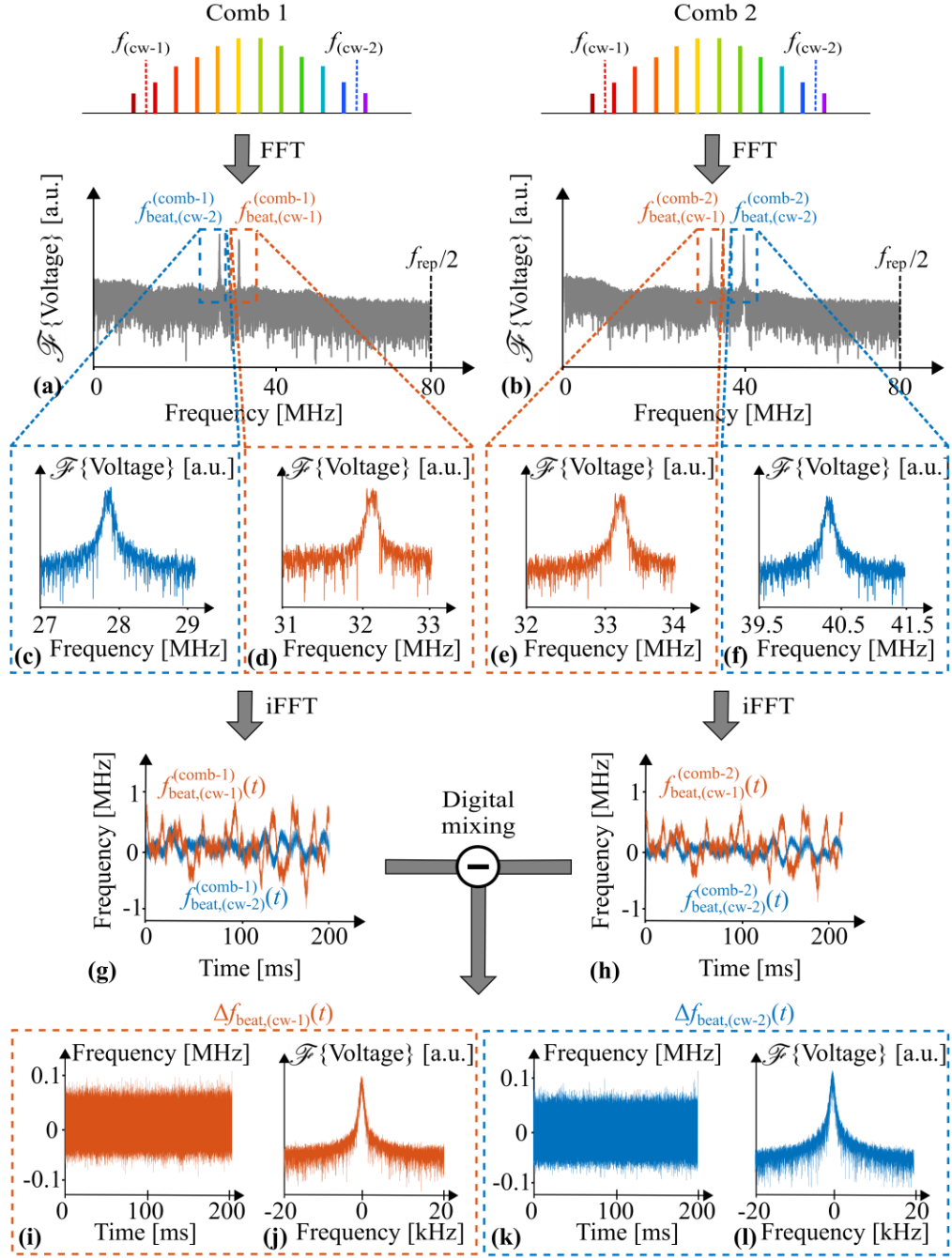

**Fig. 3.** Illustration of the analysis routine used to extract the repetition rate difference as a function of time from the measured beat note signals of the cw lasers with the two frequency combs. (a) and (b): Fourier transform of the beat note between both cw lasers with comb 1 (a); and comb 2 (b). (c)-(f) data after band-pass filtering for cw laser 1 (c) and (d); and cw laser 2 (e) and (f). (g) and (h): frequency-fluctuations of the beating between one individual cw laser with comb 1 (g); and comb 2 (h). (i)-(l): difference between the beat note signals corresponding to the same cw laser in the time-domain for cw laser 1 (i) and cw laser 2 (k); and in frequency domain for cw-laser 1 (j) and cw laser 2 (l). The data associated with $f_{(cw-1)}$ ($f_{(cw-2)}$) is shown in orange (blue). Notice that the data shown in this figure corresponds to the actual data used in the measurement of $\Delta f_{rep}(t)$ as described in more detail in section 4. The vertical axes with arbitrary units ([a.u.]) are in logarithmic scale.

## 4. Uncorrelated timing jitter of a free-running dual-comb laser

In this section, we experimentally demonstrate the performance of the proposed timing jitter measurement. For that purpose, we perform a detailed characterization of the uncorrelated repetition rate noise of a polarization-multiplexed free-running dual-comb system.

*4.1 Laser concept and modelocking characteristics*

Free-running dual-combs are a new trend in the development of dual-comb lasers [18-20, 41-50]. The idea is to design dual-comb systems with built-in passive mutual coherence which avoids the complex stabilization of two independent modelocked fs-lasers for achieving the required relative stability of the frequency combs [13-17]. The passive stability leads to low relative timing jitter between the two combs, which is important for many applications of dual-comb systems. Here we have optimized the low-noise performance of a polarization-multiplexed diode-pumped Yb:CaF$_2$ dual-comb laser system using a single-mode pump laser (schematic of the laser cavity is shown in Fig. 4(a)). Single-mode pump diodes offer high brightness at modest average power, which minimizes thermal effects and relaxes cooling requirements. Therefore, no water cooling was required for this dual-comb laser compared to our previous lasers pumped with multimode diodes [36, 49]. The suitability of single-mode pumping for dual-comb generation from a single-cavity femtosecond solid-state laser based on polarization-multiplexing was demonstrated by Kowalczyk *et al.* [50].

The dual-comb laser is based on a folded end-pumped oscillator cavity that simultaneously supports two cross-polarized laser modes. The cavity contains a semiconductor saturable absorber mirror (SESAM) [73] as an end mirror and a bulk 4.5 mm long Yb:CaF$_2$ gain crystal (4.5% doping concentration). This gain medium exhibits nearly isotropic linear-optical properties so that it provides the same gain for both polarizations inside the cavity [74]. Furthermore, it has favourable thermal properties, a broad and smooth emission spectrum and a long upper state lifetime which damps high-frequency pump intensity noise. The crystal is pumped by one single 980-nm wavelength-stabilized single-mode pump-diode (2000CHP, 3SP Technologies) that can deliver up to 850 mW output power via a single-mode fiber. To optimize the noise performance of the dual-comb system it is beneficial to use the same pump for both combs so that the pump intensity noise is the same for both intra-cavity modes. We thus split the pump into two beams of equal power which are imaged with a scale factor into the gain crystal to create two spatially separated focused pumped spots there, i.e. one for each of the two cross-polarized laser modes. The gain crystal is pumped through a flat output coupler ("OC" in Fig. 4(a)) with a transmission of 2.6% for the laser wavelength and high transmission for the pump wavelength $\lambda_{pump}$ = 980 nm. To separate the laser output beam from the pump input, we utilize a dichroic mirror.

The two cross-polarized intra-cavity beams are separated at both ends of the cavity by two 6-mm long birefringent α-BBO crystals which are cut at 45° with respect to the c-axis of the crystal. The first birefringent crystal (BC$_1$) is placed directly after the gain material while the second crystal (BC$_2$) is in front of the SESAM to yield spatially separated spots on both the gain crystal and the SESAM. The spatial separation of the modes on the SESAM is important to guarantee independent saturable absorption and to avoid saturation cross-talk between the two combs. Similarly, it is important for the two modes to be spatially separated in the gain crystal to allow for independent pumping of each mode and to avoid gain crosstalk. The two birefringent crystals are placed at the ends of the cavity to maximize the common path of the modes inside the cavity which is expected to be favourable in terms of noise properties, thereby resulting in a highly stable repetition rate difference.

Since each birefringent crystal (BC$_1$ and BC$_2$) leads to a spatial walk-off of the extraordinary wave by about 430 µm, they both introduce a delay between the ordinary and the extraordinary waves. To allow for a low repetition rate difference between the two cross-polarized combs on the order of few kHz, which is important to avoid aliasing effects in dual-comb spectroscopy, the birefringent crystals are rotated by 90° with respect to each other around the optical axis so that each polarization experiences the spatial walk-off once. This cancels the optical path length difference between the two cross-polarized beams. For fine-tuning the repetition rate difference $\Delta f_{rep}$, the second birefringent crystal BC$_2$ can be tilted to create a small path difference which results in the repetition rate difference between the two combs. The tilt of $BC_2$ does not couple to alignment of the cavity, which thus allows for a continuously tuneable repetition rate difference in the range from 20 Hz to more than 20 kHz.

To achieve fundamental soliton mode-locking [75], we balance the self-phase modulation from the gain and birefringent crystals with -1940 $fs^2$ negative group delay dispersion (GDD) introduced by the output coupler and a Gires-Tournois-Interferometer type mirror. The SESAM with a saturation fluence of 11 µJ/$cm^2$ and a modulation depth of 1.25% enabled self-starting operation and robust mode-locking of the laser.

The self-starting soliton modelocked dual-comb laser has an average output power of 130 mW per comb, a pulse duration of 115 fs, a pulse repetition rate of 160 MHz, and a small repetition rate difference of $\Delta f_{rep}$= 1.286 kHz (Fig. 4(b) - (e)). The laser is contained in a robust and compact (42 x 26 x 10 $cm^3$) prototype housing to reduce acoustical and mechanical vibrations.

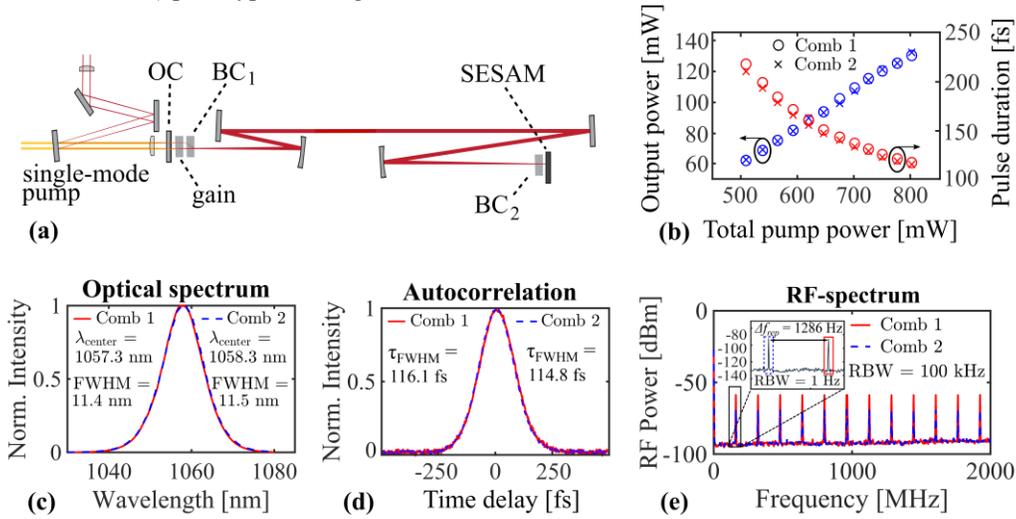

**Fig. 4.** (a) Layout of the 160 MHz dual-comb laser based on a common-path polarization-multiplexed cavity using birefringent α-BBO crystals ($BC_1$, $BC_2$) at both ends of the cavity, which gives a spatial separation in the Yb:$CaF_2$ gain crystal (CIMAP of Caen, France) and on the SESAM (OC = output coupler). (b) – (e) Characterization of the laser performance in simultaneous dual-comb lasing: (b) output power and pulse duration as a function of total pump power. The indicated pump power is split equally between the two laser modes; (c) optical spectrum for comb 1 (s-polarized) and comb 2 (p-polarized); (d) pulse duration measured with intensity autocorrelation; (e) radio-frequency (RF) spectrum with zoom on first harmonic (RBW = resolution bandwidth).

*4.2 Noise characterization of the laser*

For characterizing the timing noise of the polarization-multiplexed free-running dual-comb system in detail, we measure the phase noise of each comb individually, as well as the uncorrelated timing jitter. The measurement of the noise inherent to the repetition rate frequency ($f_{rep}$) of the individual frequency combs was carried out with a signal-source-analyzer (E5052B, Keysight) using the 3$^{rd}$ harmonic of the repetition rate. The repetition rate difference as a function of time $\Delta f_{rep}(t)$, as derived in section 3.2, is shown in Fig. 5(a). From this quantity, we can deduce the PN-PSD associated with the repetition rate difference by using Eq. (9), as discussed in section 2. The PN-PSDs of the two combs are shown in Fig. 5(b). The red and blue curve show the PN-PSD of the two individual frequency combs. The green curve shows the PN-PSD associated with the noise of the repetition rate difference $\Delta f_{rep}(t)$ and the black dashed line shows the calculated detection noise floor $S_{NF}$ of the multiheterodyne detection measurement as derived in the Appendix.

The PN-PSD of the laser suggests that the phase noise of the signal is highly correlated for the two combs as they exhibit a significant overlap within the measured frequency band. The correlation of the phase noise manifests itself in a reduction of the uncorrelated timing jitter between the two combs by about 14 dB compared to the noise in $f_{rep}$ of a single comb (Fig. 5(b)). This is equivalent to a reduction of the relative rms timing jitter in a frequency band from [100 Hz, 1 MHz] from 58 fs (for comb 1) and

65 fs (for comb 2) to only 13 fs for the uncorrelated part (Fig. 5(c)) as can be found from Eq. (11). Notice that the data shown in Fig. *3* corresponds to the actual data used for finding the results displayed in Fig. 5.

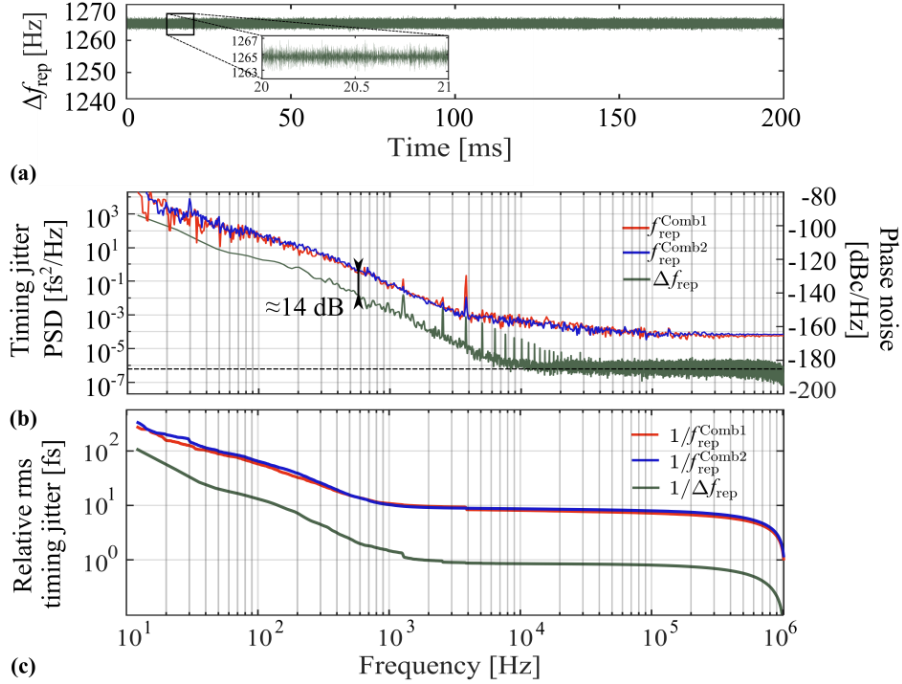

**Fig. 5.** (a) $\Delta f_{rep}$ as a function of time, measured with the multiheterodyne detection setup. (b) Phase noise power spectral density (PN-PSD) for the individual frequency combs and the uncorrelated timing jitter between the two combs — deduced from the trace shown in (a). The measurement of the $f_{rep}$ phase noise was carried out with a signal-source-analyzer (E5052B, Keysight) using the 3$^{rd}$ harmonic of the repetition rate. The measurements with the signal-source-analyzer (SSA) have been averaged over five measurements, while the uncorrelated timing jitter results from a single measurement with the multiheterodyne beat note detection scheme. The decrease of the noise floor at 1 MHz is the result of a digital bandpass filter that was applied in the post-processing. (c) relative rms timing jitter.

The peaks in the $f_{rep}$ phase noise and the uncorrelated timing jitter (see Fig. 5(b)) at integer multiples of the repetition rate difference originate from cross-talk between the two polarization multiplexed combs inside the laser cavity due to the common path. This manifests itself in peaks at integer multiples of the repetition rate for the uncorrelated timing jitter, and in peaks at integer multiples of $3 \cdot \Delta f_{rep}$ for the $f_{rep}$ phase noise because this measurement was carried out with a signal-source-analyzer (SSA; E5052B, Keysight) using the 3$^{rd}$ harmonic of the repetition rate.

The measured PN-PSD of the two individual frequency combs reaches a detection noise floor of $6.7 \times 10^{-4}$ fs$^2$/Hz at 5 kHz, which corresponds to the frequency dependent phase noise sensitivity limit of the SSA. At higher offset frequencies from the carrier, the measurement is still limited by the detection noise floor of the SSA. The PN-PSD of the repetition rate difference $\Delta f_{rep}(t)$ reaches a timing-jitter detection noise floor of around $8.0 \times 10^{-7}$ fs$^2$/Hz at 10 kHz. This agrees well with the calculated detection noise floor $S_{NF}$ of the multiheterodyne detection measurement at $5.8 \times 10^{-7}$ fs$^2$/Hz as derived in the Appendix.

It is useful to understand the origin of this noise floor. The relative intensity noise (RIN) of the detected beat note signals is dominated by the RIN of the cw lasers $S_{cw}^{RIN}$, which we measured to be $S_{cw}^{RIN} = -155$ dBc/Hz using the same SSA as mentioned above. Since the signal corresponds to a beat note between the cw laser and a single comb line, we can estimate its contribution to the PN-PSD noise floor as

$$S_{\Delta f_{\text{rep}}}^{\text{PN}} = \frac{S_{\text{cw}}^{\text{RIN}} \cdot P_{\text{cw}}^{\text{optical}}}{P_{\text{SCL}}^{\text{optical}} \cdot \Delta N^2} \bigg/ \left(2\pi f_{\text{rep}}\right)^2, \tag{14}$$

in units of s²/Hz where $P_{\text{cw}}^{\text{optical}}$ is the optical power of the used cw lasers ($P_{\text{cw}}^{\text{optical}} = 18$ mW), $P_{\text{SCL}}^{\text{optical}}$ is the power of the single comb line that is beating with the cw laser ($P_{\text{SCL}}^{\text{optical}} = 6.7$ nW) and $\Delta N$ is the number of comb lines between the two cw lasers ($\Delta N \approx 53'187$). A detailed derivation of Eq. (14) is given in section 7.

From Eq. (14) we learn that as the optical power in the frequency combs $P_{\text{comb}}$ is increased, the power of the single comb line $P_{\text{SCL}}^{\text{optical}}$ that is beating with the cw laser grows as well, so that consequently the noise floor of the measurement is reduced. However, there is a limit to how much power can be sent onto the photodetector before it starts to saturate. To bypass this limitation, it is important to realize that only two very narrow spectral bands of the frequency combs optical spectra around the frequency of the two cw lasers contribute to the heterodyne signals that are relevant for the technique. Consequently, it is sufficient to send only those narrow spectral bands onto the detector, which enables a higher spectral density at the same average power on the photodiode. We implemented this spectral filtering with a 300 µm thick Fabry-Perot etalon (LightMachinery Inc). Both frequency combs are directed through the etalon in a non-collinear configuration. By rotating the etalon, the angle of incidence for both combs can be adjusted until the transmission peaks in the optical spectra of both combs perfectly overlap. In a final step one then needs to tune the frequency of the cw-lasers to one of the transmission peaks. This allows to significantly lower the noise floor of the measurement: The response of the 0.8-mm² InGaAs photodiode that we used for the measurement remains linear to pulses even with average powers above 15 mW in the case of 50-Ω termination. By sending the light of the dual-comb system through the etalon prior to detection, only about 2.5 mW arrive at the photodiode, which is far from the saturation limit. In fact, if the dual-comb laser had a higher output power of around 800 mW per comb, we could send a total power of 15 mW onto the detector to potentially reduce the noise floor down to below $9.6 \times 10^{-8}$ fs²/Hz. Combining a thinner etalon, broad optical spectra, and low-noise cw lasers, a noise floor of $1 \times 10^{-8}$ fs²/Hz (= 100 zs/$\sqrt{Hz}$) is well within reach for this measurement technique.

Notice that the quantification of the timing noise does not directly determine the linewidth of the individual comb lines since these are also affected by noise in $\Delta f_{\text{CEO}}(t)$. As mentioned in section 2, the proposed measurement technique allows to determine the carrier envelope offset frequency difference as a function of time, provided that the comb line index is known. Alternatively, the intermediate digital mixing product from Eq. (3) can be used directly to calculate the FWHM linewidth of a single comb tooth via the β-separation-line approximation [66]. We found this to be 515.8 Hz for a lower cut-off frequency of 5 Hz.

## 5. Comparison with other noise measurement techniques

In this section, we compare the timing jitter inferred from the proposed multiheterodyne measurement technique with the timing jitter extracted from interferograms. From the discussion in section 2 we know that the PN-PSD associated with $\Delta f_{\text{rep}}(t)$ can be used to infer the timing jitter between the two pulse trains. However, the information about the timing of the two pulse trains is also imprinted on the interferograms, which are measured alongside the beat note signals to determine the comb line index difference as discussed in section 2. As the interferograms repeat every $1/\Delta f_{\text{rep}}$, the relative timing jitter between the time $t_n$ of the $n$-th interferogram and its expected arrival time of $n \cdot T$ is given by

$$\tau^{\text{rms}}[n] = t_n - n \cdot T, \tag{15}$$

where $T = 1/\langle \Delta f_{\text{rep}} \rangle$ is the average period of the signal. This jitter is visualized in Fig. 6(a). Similarly, the period timing jitter is defined as the difference between the $n$-th period of the signal and the average period $T$

$$\tau^{\text{period}}[n] = (t_n - t_{n-1}) - T. \tag{16}$$

The sampling rate of the interferogram measurement is limited by $\Delta f_{\text{rep}}$, which in this case is at 1.286 kHz, whereas the multiheterodyne beat note detection scheme allows measuring the repetition rate difference with a sampling frequency that goes up to more than 1 MHz. The higher sampling rate makes the latter the superior technique for measuring the timing jitter in terms of performance. However, the approach based on the interferograms is simpler and it can still be used as a cross-check of the multiheterodyne beat note detection scheme. For that purpose, we multiply the timing jitter of the interferograms by the comb factor $\Delta f_{\text{rep}}/f_{\text{rep}}$ to relate the RF-domain to the optical domain where the timing jitter describes the stability of the optical pulses.

We compute the rms of the relative and period timing jitter associated with the interferograms given by Eqs. (15) and (16), multiply this with the comb factor and compare it to the timing jitter inferred from the PN-PSD obtained from the multiheterodyne beat note measurement. The result of this comparison is shown in Fig. 6(b). The relative rms timing jitter approaches 283.4 fs when integrating from 1 MHz down to 5 Hz which agrees well with the relative rms timing jitter of 282.1 fs inferred from the interferogram peaks. Similarly, for the period timing jitter, the beat note measurement predicts a relative rms timing jitter of 15.6 fs [5 Hz, 1 MHz] while the interferogram peaks suggest 16.9 fs. Those results indicate that the relative rms and period timing jitter inferred from the two different measurement techniques agree well. Furthermore, Fig. 6(b) illustrates the benefits of the high sampling rate inherent to the approach based on optical heterodyne detection: Instead of just a single value that indicates the timing jitter associated with a dual-comb – as it is the case for the interferogram measurement – the multiheterodyne beat note measurement provides information about the frequency-dependence of the timing jitter, including the influence of noise frequencies above $\Delta f_{\text{rep}}$. These measurements are valuable for optimizing the noise performance of dual-comb systems by providing detailed information about their noise properties. This information can also be used to study the noise characteristics of dual-comb systems: We observe for example that the relative rms timing jitter diverges more rapidly than the period timing jitter for low frequencies. This behavior is a consequence of the definition of the period jitter which involves sampling at a rate of $\Delta f_{\text{rep}}$. Hence, this jitter cannot diverge rapidly unless $\Delta f_{\text{rep}}$ varies significantly over a timescale of $1/\Delta f_{\text{rep}}$. In contrast, the relative rms timing jitter involves the accumulated delay between the pulses relative to an ideal pulse train. This quantity can become arbitrarily large if one waits long enough (similar to a random walk).

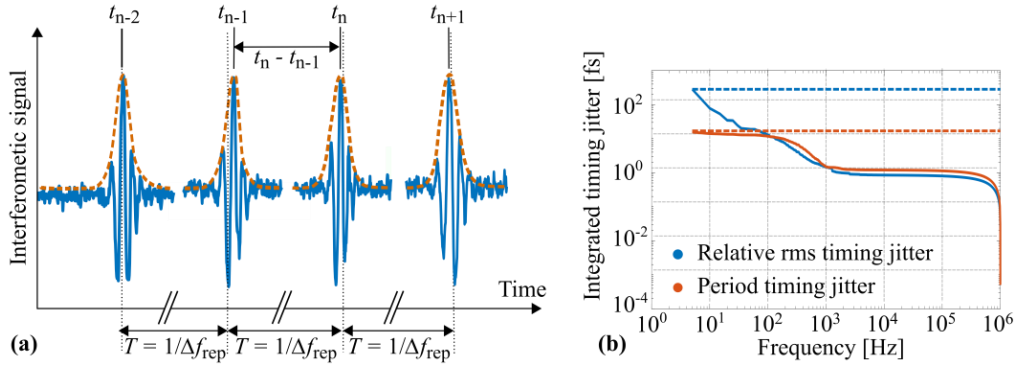

**Fig. 6.** (a) Interferogram peaks to illustrate the concept of relative rms and period timing jitter in interferometric measurements. Dashed vertical lines indicate the average arrival time defined by the mean of the inverse repetition rate difference of the two pulse trains. Solid vertical lines indicate the peak of the individual interferograms. (b) Comparison of integrated timing jitter obtained from the interferogram measurement (dashed line) and multiheterodyne beat note measurement (solid line).

An alternative to characterizing the timing jitter via the interferogram measurement is to use a frequency counter [76]. However, it exhibits similar limitations as the interferogram measurement meaning that its sampling rate cannot be higher than the repetition rate difference. The restricted sampling rate heavily limits the applicability of this measurement for dual-comb systems with low repetition rate difference.

## 6. Conclusion

The optical multiheterodyne detection approach for phase noise measurements demonstrated in this paper allows for the precise characterization of the relative timing jitter of two pulse trains. This technique is valuable for a wide range of dual-comb systems as it can be applied whenever the repetition rate difference is small compared to the repetition rate. Specifically, the measurement is not restricted to dual-comb systems with active stabilization, since it is no requirement that the delay between the two combs stays small. This makes the measurement technique particularly suitable for free-running dual-comb lasers. The detection noise floor at $8.0\times10^{-7}$ fs$^2$/Hz provides a high sensitivity of the measurement suitable for ultra-low noise dual comb systems. The sensitivity of the measurement can be further increased by improving the filtering of the optical spectrum at the frequencies of the two cw lasers or by sending more power on the photodetector. In addition, one could also nonlinearly broaden the optical spectrum via self-phase modulation to increase the spectral separation of the cw lasers which consequently leads to a larger number of comb lines between the two cw lasers.

To experimentally demonstrate the proposed method, we utilized it for characterizing the noise performance of a novel polarization-multiplexed single-cavity dual-comb solid-state laser with a low repetition rate difference. The laser operates at 160 MHz repetition rate with 115 fs pulse duration and 130 mW average power per comb. We measured a relative rms timing jitter of 58 fs for Comb 1 and 65 fs for Comb 2 when integrating from 100 Hz to 1 MHz. Due to the passive stability of the single-cavity dual-comb laser, the relative timing jitter between the combs is suppressed by about a factor of 5 compared to the relative rms timing jitter of the individual combs, resulting in a timing jitter of around 13 fs [100 Hz, 1 MHz]. We confirmed this result and the proposed measurement technique itself with a comparison to a timing jitter measurement based on an interferogram measurement. Due to this low relative timing jitter, the single-mode-diode pumped single-cavity dual comb laser is highly compatible with practical dual-comb spectroscopy, ranging, and sampling applications for which a typical measurement duration is in the millisecond regime or even faster.

## 7. Appendix

*7.1 Derivation of the noise floor for the optical heterodyne detection approach*

The relative intensity noise (RIN) of each laser contributes to noise in the measured photocurrent. At the relevant frequencies for the beat notes (i.e. in the megahertz range), the modelocked laser noise contribution is dominated by shot noise. In contrast, the cw lasers have RIN above the shot noise limit, which we measured to be white noise with a one-sided power spectral density $S_{cw}^{RIN}$ = -155 dBc/Hz at 10 MHz by using our signal source analyzer. Hence, for these measurements, the photocurrent noise is dominated by the RIN of the cw lasers. In the following, we express the RIN of the cw lasers in units of reciprocal Hertz as $S_{cw}^{RIN} = 2\cdot 10^{-\frac{155}{10}}$ Hz$^{-1}$.

Since the photodetectors produce a photocurrent that is proportional to the optical power, the power of the signal detected by the signal source analyzer is proportional to the square of the optical power. The one-sided electronic power spectral density for the cw-laser is thus proportional to

$$S_{cw}^{electronic} = S_{cw}^{RIN} \cdot \left(P_{cw}^{optical}\right)^2, \qquad (17)$$

where $P_{cw}^{optical}$ is the optical power of the used cw laser ($P_{cw}^{optical}$ = 18 mW). The RIN associated with the multiheterodyne detection approach can then be inferred from $S_{cw}^{electronic}$ by dividing this expression by the optical power of the detected signal squared (the signal corresponds to the beat note of a cw laser with one single comb line which is closest to the cw laser). The effective optical power of this signal is given by

$$P_{\text{signal}}^{\text{optical}} = \sqrt{P_{\text{cw}}^{\text{optical}}} \cdot \sqrt{P_{\text{SCL}}^{\text{optical}}}, \tag{18}$$

where $P_{\text{SCL}}^{\text{optical}}$ is the optical power of the single comb line that is beating with the cw laser. The power of this comb line can be estimated from the optical spectra and power associated with the two frequency combs (see Fig. 4) as $P_{\text{SCL}}^{\text{optical}} = 6.7$ nW. Consequently, the RIN associated with the multiheterodyne detection approach is given by

$$S_{\text{heterodyne}}^{\text{RIN}} = \frac{S_{\text{cw}}^{\text{electronic}}}{\left(P_{\text{signal}}^{\text{optical}}\right)^2} = \frac{S_{\text{cw}}^{\text{RIN}} \cdot P_{\text{cw}}^{\text{optical}}}{P_{\text{SCL}}^{\text{optical}}}. \tag{19}$$

This expression is a theoretical prediction for the intensity noise of the detected beat note signal. For the beat note between a single comb line and a cw laser, the white noise of the cw lasers at megahertz frequencies contributes equally to the phase and amplitude quadrature of the measured microwave signal [77], i.e. the noise is equally distributed to amplitude and phase. Consequently, the theoretical prediction for the phase noise (PN) of the detected beat note signal may be written as

$$S_{\text{heterodyne}}^{\text{PN}} = \frac{S_{\text{cw}}^{\text{RIN}} \cdot P_{\text{cw}}^{\text{optical}}}{P_{\text{SCL}}^{\text{optical}}}. \tag{20}$$

Finally, in the heterodyne detection approach the noise fluctuations of the optical beat note signals were intentionally increased by the number of comb lines between the two cw lasers $\Delta N$ to improve the sensitivity of the measurement ($\Delta N \approx 53'187$). To account for this in the theoretical expression, we need to divide $S_{\text{heterodyne}}^{\text{PN}}$ by $\Delta N^2$ which yields

$$S_{\Delta f_{\text{rep}}}^{\text{PN}} = \frac{S_{\text{cw}}^{\text{RIN}} \cdot P_{\text{cw}}^{\text{optical}}}{P_{\text{SCL}}^{\text{optical}} \cdot \Delta N^2}. \tag{21}$$

Since $S_{\text{cw}}^{\text{RIN}}$ describes the RIN of the cw lasers in Hz$^{-1}$, the noise floor in Eq. (21) is also expressed in units of Hz$^{-1}$. However, we can describe the noise floor also in s$^2$/Hz as

$$S_{\Delta f_{\text{rep}}}^{\text{PN}} = \frac{S_{\text{cw}}^{\text{RIN}} \cdot P_{\text{cw}}^{\text{optical}}}{P_{\text{SCL}}^{\text{optical}} \cdot \Delta N^2} \bigg/ \left(2\pi f_{\text{rep}}\right)^2, \tag{22}$$

which corresponds to the expression in Eq. (14).

**Funding.** H2020 European Research Council (966718); Schweizerischer Nationalfonds zur Förderung der Wissenschaftlichen Forschung (40B2-0_180933); Innosuisse - Schweizerische Agentur für Innovationsförderung (40B2-0_180933).

**Acknowledgments.** We thank Dr. Valentin Wittwer and Prof. Dr. Thomas Südmeyer from Université de Neuchâtel for lending the Toptica CTL 1050. We also thank our collaborator Prof. Dr. Patrice Camy, CIMAP of Caen, France for manufacturing the Yb:CaF$_2$ gain crystals.

**Disclosures.** The authors declare no conflicts of interest.

**Data availability.** Data underlying the results presented in this paper is available at ETH Zurich Research Collection library [78].